\documentclass[journal]{IEEEtran}
\usepackage{amsmath,amsfonts,graphicx,tabularx,color}
\usepackage{amssymb,textcomp,mathtools, amsthm}
\usepackage{bm,upgreek,algorithm,algpseudocode,hyperref}
\usepackage{multirow,booktabs,hhline,array}
\usepackage{cite,url,makecell,setspace,listings}
\RequirePackage{doi}

\definecolor{codegreen}{rgb}{0,0.6,0}
\definecolor{codegray}{rgb}{0.5,0.5,0.5}
\definecolor{codepurple}{rgb}{0.58,0,0.82}
\definecolor{backcolour}{rgb}{0.95,0.95,0.92}

\lstdefinestyle{mystyle}{
    backgroundcolor=\color{backcolour},   
    commentstyle=\color{codegreen},
    keywordstyle=\color{magenta},
    numberstyle=\tiny\color{codegray},
    stringstyle=\color{codepurple},
    basicstyle=\ttfamily\scriptsize,
    breakatwhitespace=false,         
    breaklines=true,                 
    captionpos=b,                    
    keepspaces=true,                 
    numbers=left,                    
    numbersep=5pt,                  
    showspaces=false,                
    showstringspaces=false,
    showtabs=false,                  
    tabsize=2
}

\renewcommand{\vec}[1]{\bm{\mathrm{#1}}}

\theoremstyle{definition}

\theoremstyle{plain}

\theoremstyle{remark}

\title{Music Source Separation with Band-split RNN}

\author{Yi~Luo, Jianwei~Yu}

\begin{document}
%\onecolumn
%\doublespacing
\maketitle
\setlength{\abovedisplayskip}{2pt}
\setlength{\belowdisplayskip}{2pt}
\setlength{\abovedisplayshortskip}{2pt}
\setlength{\belowdisplayshortskip}{2pt}

\begin{abstract}
The performance of music source separation (MSS) models has been greatly improved in recent years thanks to the development of novel neural network architectures and training pipelines. However, recent model designs for MSS were mainly motivated by other audio processing tasks or other research fields, while the intrinsic characteristics and patterns of the music signals were not fully discovered. In this paper, we propose band-split RNN (BSRNN), a frequency-domain model that explictly splits the spectrogram of the mixture into subbands and perform interleaved band-level and sequence-level modeling. The choices of the bandwidths of the subbands can be determined by a priori knowledge or expert knowledge on the characteristics of the target source in order to optimize the performance on a certain type of target musical instrument. To better make use of unlabeled data, we also describe a semi-supervised model finetuning pipeline that can further improve the performance of the model. Experiment results show that BSRNN trained only on MUSDB18-HQ dataset significantly outperforms several top-ranking models in Music Demixing (MDX) Challenge 2021, and the semi-supervised finetuning stage further improves the performance on all four instrument tracks.
\end{abstract}

\begin{IEEEkeywords}
Music separation, Neural network, Deep learning
\end{IEEEkeywords}

\section{Introduction}
\label{sec:intro}
The task of music source separation (MSS) has drawn more and more attention in the community due to its wide application in music remixing \cite{gillet2005extraction, woodruff2006remixing, pons2016remixing}, music information retrieval (MIR) \cite{ono2010harmonic, mesaros2010automatic, itoyama2011query, rosner2014classification, lin2021unified}, and music education \cite{dittmar2012music, cano2014pitch}. The advances in MSS models can also shed light on the investigation of novel model designs for other related tasks such as speech enhancement and speech separation \cite{macartney2018improved, giri2019attention, defossez2020real, jenrungrot2020cone, liu2021voicefixer}. Moreover, since music signals are typically recorded at a higher sample rate than signals in telecommunication systems (e.g., 44.1k Hz) and have been artificially edited or manipulated by professionals, MSS is in general more challenging than narrow-band or wide-band speech enhancement and separation tasks. Developing high-quality MSS models is thus important and necessary towards building a robust universal source separation system in diverse and complicated real-world scenarios.

Modern MSS models are generally built upon deeper and more complicated neural network architectures. While frequency-domain systems are the mainstream \cite{nugraha2016multichannel, jansson2017singing, luo2017deep, chandna2017monoaural, park2018music, takahashi2018mmdenselstm, takahashi2020d3net, hennequin2020spleeter, li2021sams, kong2021decoupling, liu2021cws}, recent works have also focused on time-domain systems \cite{stoller2018wave, defossez2019music, defossez2019demucs, samuel2020meta} as well as the fusion of time-domain and frequency-domain systems \cite{defossez2021hybrid, kim2021kuielab}. However, many of the MSS models were motivated by existing system architectures in other research fields. For example, models in speech separation can be directly applied to MSS without modifications \cite{huang2014deep, huang2015joint, hershey2016deep, luo2017deep, luo2019conv, samuel2020meta}, and architectures in image segmentation \cite{ronneberger2015u}, human pose estimation \cite{newell2016stacked} and image recognition \cite{huang2017densely} have been directly utilized in various recent MSS models. Although many of those prior arts have proven effective in terms of the separation performance, the question of why they are effective in music data and how they can be modified to better explore the intrinsic characteristics of music signals is not easy to answer. Moreover, since existing works in speech enhancement and separation typically only considers narrow-band and wide-band signals, how to adjust them to super wide-band music signals is also worth investigating. The characteristics of singing voice and speech can also be different in terms of fundamental frequency, loudness and formants \cite{livingstone2013acoustic}, and speech separation models may have the potential to be further improved on singing voice if such properties could be considered in the modification of the models.

In this paper, we propose \textit{band-split RNN (BSRNN)}, a frequency-domain MSS model that is specially designed for high sample rate signals with flexible and explicit segregation and modeling of different frequency bands. BSRNN splits a spectrogram into a series of subband spectrograms with a set of predefined bandwidths, and the bandwidths are adjusted for different instrument types accordingly. The subband spectrograms are then transformed to generate a series of features with a same feature dimension, and stacked residual recurrent neural network (RNN) layers are utilized to perform interleaved cross-band and cross-sequence modeling similar to dual-path networks for speech separation \cite{luo2020dual}. Each subband feature in the output of the last residual RNN layer is then transformed by a multilayer perceptron (MLP) to generate its corresponding complex-valued time-frequency mask, and the mask is applied to the corresponding subband mixture spectrogram to generate the estimated target source spectrogram. All estimated subband spectrograms are then concatenated to form the full spectrogram of the target source. The most important module in BSRNN is the band-splitting module, which enable us to incorporate prior knowledge on the target source into the model design. For example, if we know in advance that the target source mainly lies in lower frequency parts with a relatively low fundamental frequency, we can perform fine-grained band-splitting scheme at lower frequency parts to increase the frequency resolution and coarse-grained band-splitting scheme at higher frequency parts to save the computational complexity. We show by experiments that band-splitting schemes do play an important role in the separation performance, and different musical instruments do have their own band-splitting scheme to obtain a performance gain. By properly selecting the band-splitting bandwidths, BSRNN trained only on MUSDB18-HQ dataset \cite{musdb18-hq} is able to significantly outperform top-ranking models in Music Demixing (MDX) Challenge 2021.

It is well-known that the performance and generalization ability of many MSS models can be limited by the size of available high-quality clean training data. Several existing works have proposed methods for digging valid segments from unlabeled data by a source activity detector and using them for data augmentation purpose during training \cite{defossez2019demucs, wang2021semi}, and several other works attempted to perform separation on the unlabeled data with a pre-trained model to generate pseudo labels \cite{tzinis2022remixit, tzinis2022continual}. We also describe a semi-supervised finetuning pipeline that can be viewed as a mixed pipeline of recent works, which bypasses the requirement of a separately trained source activity detector by directly using a strong pre-trained model as both its own source activity detector and pseudo-label generator. A self-boosting scheme is applied to gradually improve the quality of the generated pseudo-label signals by continuously replacing the pre-trained model by the new best model found in the finetuning stage. This setting allows us to use both the clean and noisy signals in the unlabeled data in the finetuning stage. Experiments show that the semi-supervised finetuning pipeline can further improve the performance of the model in all instrument tracks.

The rest of the paper is organized as follows. Section~\ref{sec:BSRNN} introduces the BSRNN architecture. Section~\ref{sec:semi} describes the semi-supervised data finetuning pipeline. Section~\ref{sec:config} provides the detailed configurations for training and evaluation. Section~\ref{sec:result} presents the experiment results and analysis. Section~\ref{sec:conclusion} concludes the paper.

\section{Band-split RNN}
\label{sec:BSRNN}
\begin{figure*}[!ht]
	\small
	\centering
	\includegraphics[width=2\columnwidth]{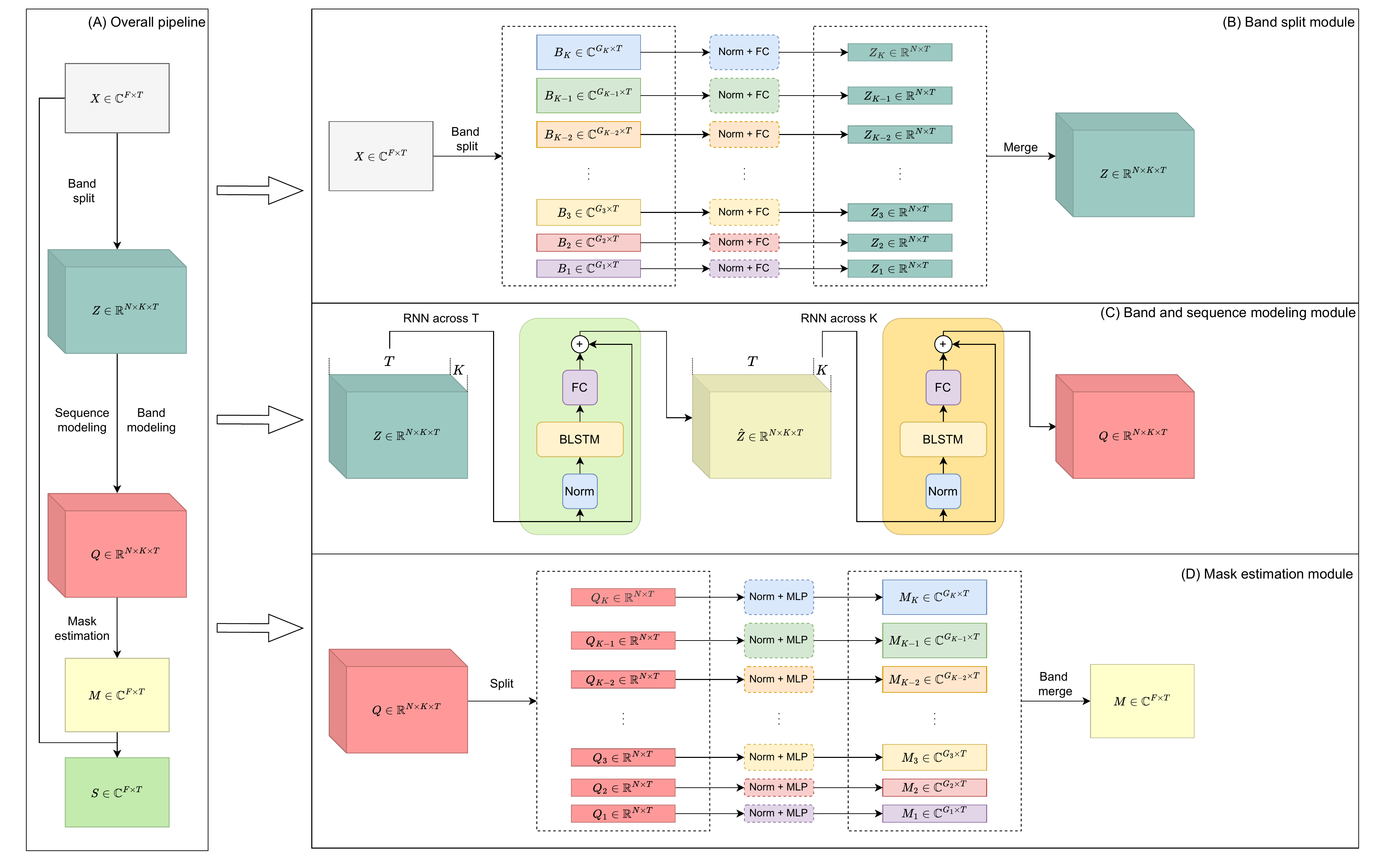}
\caption{(A) The overall pipeline for the BSRNN model, which consists of a band split module, a sequence and band modeling module, and a mask estimation module. (B) The design of the band split module. (C) The design of the sequence and band modeling module. (D) The design of the mask estimation module.}
\label{fig:BSRNN}
\end{figure*}

Figure~\ref{fig:BSRNN} (A) shows the overall pipeline of the BSRNN model, which contains a \textit{band split} module, a \textit{band and sequence modeling module}, and a \textit{mask estimation} module.

\subsection{Band Split Module}

Figure~\ref{fig:BSRNN} (B) shows the design of the band split module. The module takes the complex-valued spectrogram $\vec{X} \in \mathbb{C}^{F\times T}$ generated by short-time Fourier transform (STFT) as input, where $F$ and $T$ are the frequency and temporal dimensions, respectively, and split it into $K$ subband spectrograms $\vec{B}_i \in \mathbb{C}^{G_i\times T}, i=1,\ldots, K$ with predefined bandwidth $\{G_i\}_{i=1}^K$ satisfying $\sum_{i=1}^K G_i = F$. The real and imaginary part of each subband spectrogram $\vec{B}_i$ is then concatenated and passed to a layer normalization module \cite{ba2016layer} and a fully-connected (FC) layer to generated a real-valued subband feature $\vec{Z}_i \in \mathbb{R}^{N\times T}$. Note that since $\{G_i\}_{i=1}^K$ can all be different, each subband spectrogram has its own normalization module and FC layer. All $K$ subband features $\{\vec{Z}_i\}_{i=1}^K$ are then merged to generate a transformed fullband feature tensor $\vec{Z} \in \mathbb{R}^{N\times K \times T}$.

\subsection{Band and Sequence Modeling Module}

Figure~\ref{fig:BSRNN} (C) shows the band and sequence modeling module. Similar to the dual-path RNN architecture \cite{luo2020dual}, BSRNN performs interleaved sequence-level and band-level modeling via two different residual RNN layers. The sequence-level RNN is first applied to $\vec{Z}$ across the temporal dimension $T$, where the $K$ subband features share a same RNN since they have the same feature dimension $N$. This is to save the model size and allow parallel processing across subbands. The band-level RNN is then applied to $\vec{Z}$ across the band dimension $K$, where the RNN is assumed to capture the intra-band feature dependencies across the $K$ subbands at each frame. Both sequence-level and band-level RNNs share a same design, where a group normalization module is first applied to the input of the module, and then a BLSTM layer followed by an FC layer is applied to perform the actual modeling. Residual connection is added between the input and the output of the FC layer. Multiple such RNNs can be stacked to create a deeper architecture, and the output of the last layer is denoted by $\vec{Q} \in \mathbb{R}^{N\times K \times T}$.

\subsection{Mask Estimation Module}

The mask estimation module calculates a complex-valued time-frequency (T-F) mask to extract the target source. $\vec{Q}$ is first split into $K$ features $\{\vec{Q}_i\}_{i=1}^K \in \mathbb{R}^{N\times T}$ where each feature corresponds to the transformed feature for a subband, and each subband feature is passed to a layer normalization module followed by a multilayer perceptron (MLP) with one hidden layer to generate the real and imaginary parts of the T-F masks $\vec{M}_i \in \mathbb{C}^{G_i\times T}, i=1,\ldots, K$. The use of MLP follows the observation in \cite{li2022on} where it was reported that a simple MLP can effectively estimate better T-F masks compared to a plain FC layer. Similar to the band split module, each subband feature has its own normalization module and MLP. All $\vec{M}_i$ are then merged into the fullband T-F mask $\vec{M} \in \mathbb{C}^{F\times T}$ and multiplied with $\vec{X}$ to generate the target spectrogram $\vec{S} \in \mathbb{C}^{F\times T}$.

\subsection{Discussion}

It is easy to observe that BSRNN can be connected to recent works on dual-path and multi-path networks \cite{kinoshita2020multi, chen2020dual, subakan2021attention, dang2022dpt, pandey2022tparn}, group-splitting modules \cite{luo2021ultra, luo2021group}, and super wideband models \cite{zhang2022two, lv2022s}. Most dual-path architectures split a sequential feature into chunks and perform interleaved local and global processing, and the additional paths in multi-path models were proposed to either split the sequential feature into finer-scale chunks or to apply on extra dimensions such as the spatial dimension in multi-channel signals. However, as the dual-path architecture was originally proposed for time-domain separation systems with relatively small window and hop size \cite{luo2020dual}, its necessity and importance become less crucial in the successful optimization of the model. We also empirically find that replacing the plain BLSTM by dual-path RNN for the sequence modeling module does not lead to a performance gain in BSRNN.

Group-splitting modules were mainly proposed to build lightweight models with fewer model parameters and computational cost, where the main idea was to split a feature vector into groups and perform group-level sequential modeling and intra-group dependency modeling \cite{luo2021ultra, luo2021group}. Such group splitting and communication scheme was originally proposed for time-domain systems where the feature vectors do not have a clear frequency-dependent pattern, hence the intra-group dependency modeling module was designed to ignore the sequential order of the grouped sub-features. Such group-splitting scheme might not be as effective as in speech separation, since different instruments may have significantly different frequency range and timbres and an explicit frequency-dependent feature extraction scheme can be helpful. This serves as the main motivation for us to explicitly split the frequency components to subband features and use a sequential-order-sensitive module to capture the intra-band dependencies. Similar to the group communication models in time-domain systems, we still share the sequence modeling layer across all subbands, as it allows parallel processing across the subbands, saves the model size, and empirically leads to better performance compared to using a separate layer for each subband.

Current models for super wideband speech enhancement split the frequency componenets into low-frequency and high-frequency bands and use either parallel or sequential network building blocks to process them \cite{zhang2022two, lv2022s}. Such band-splitting schemes are simple and coarse, and the intra-band dependencies are not explicitly modeled. BSRNN performs a fine-grained band-splitting scheme that attempts to cover more detailed harmonic patterns in the music signals. On the other hand, BSRNN does not perform strict frequency-bin-level modeling as prior works on automatic speech recognition \cite{li2015lstm, li2016exploring} and fullband speech enhancement \cite{hao2021fullsubnet, chen2022fullsubnet+} in order to save the model complexity, memory footprint and processing speed.

\section{Semi-supervised Finetuning Pipeline}
\label{sec:semi}
Collecting high-quality realistic training data with all clean target sources is challenging for not only music signals but a wide range of other types of signals such as environmental sounds. More importantly, as the choice of musical instruments and their arrangements can be completely different in different songs, it is hard to collect clean sources for all possible musical instruments with a satisfying quality. However, a model trained with a limited amount of labeled data may fail in songs with different genres and choices of musical instruments. In this section, we describe how we finetune the model trained with labeled data on additional unlabeled data with semi-supervised data sampling.

\begin{figure}[!ht]
	\small
	\centering
	\includegraphics[width=0.6\columnwidth]{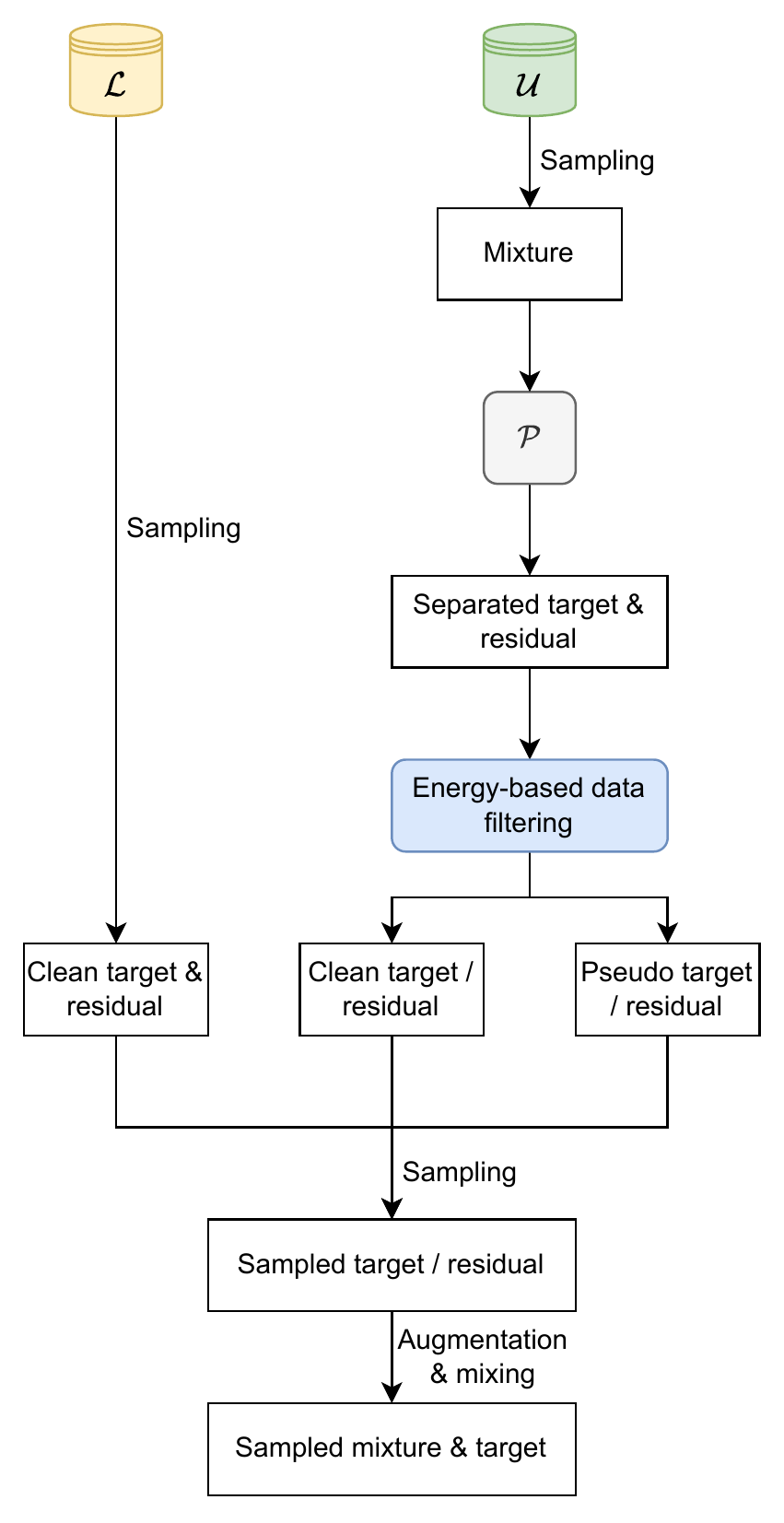}
\caption{The semi-supervised data sampling pipeline for model finetuning.}
\label{fig:semi}
\end{figure}

\subsection{Semi-supervised Data Sampling}
\label{sec:sampling}
Given a model $\mathcal{P}$ trained on a small-scale labeled dataset $\mathcal{L}$ and a large-scale unlabeled dataset $\mathcal{U}$, we generate new training samples by the sampling pipeline described in Figure~\ref{fig:semi}. The core concept of our semi-supervised data sampling pipeline is that we treat the pre-trained model as both a pseudo-label generator \cite{wang2021semi} and a source activity detector \cite{kumar2018music}. We sample a segment of clean target and residual signals from $\mathcal{L}$ as the supervised training pipeline, where the ``residual'' signal is defined as the signal that does not contain the target source (e.g., accompaniment when vocal track is the target). We also sample a segment of mixture signal from $\mathcal{U}$ and pass it to the pre-trained model $\mathcal{P}$ to generate the separated target and residual signals, where the residual signal is defined by subtracting the separated target signal from the mixture. We then use an energy-based data filtering method to detect whether the separated signals are clean or noisy/distorted:
\begin{enumerate}
    \item If the energy difference between the mixture and separated target signals, measured on decibel scale, is larger than 30 dB, then the mixture segment is treated as a clean residual segment.
    \item If the energy difference between the mixture and separated residual signals, measured on decibel scale, is larger than 30 dB, then the mixture segment is treated as a clean target segment.
\end{enumerate}
This simple energy-based method directly used the pre-trained model as the target source activity detector. The separated signals are not used when the mixture signal is treated as a clean target or residual segment, and they are treated as \textit{pseudo labels} when the mixture signal contains both the target and residual signals. All clean and pseudo signals are then gathered and one target and one residual signal are resampled from them. Optional data augmentation and mixing process can be applied to the signals, and then the transformed target and residual signals are summed to generate the mixture signal used for the finetuning stage.

During the finetuning stage, we define a new model $\mathcal{Q}$ initialize it by $\mathcal{P}$ and train it on samples generated by the aforementioned data sampling pipeline. The pre-trained model $\mathcal{P}$ can be fixed or sequentially updated by the new $\mathcal{Q}$ \cite{tzinis2022remixit}, and we replace $\mathcal{P}$ by the new $\mathcal{Q}$ whenever $\mathcal{Q}$ achieves a better performance than $\mathcal{P}$ on the validation set.

\subsection{Discussion}

The data sampling pipeline described here can be connected to a wide range of existing works on knowledge distillation \cite{hinton2015distilling, asami2017domain, xu2020knowledge, gou2021knowledge}, pseudo label generation \cite{lee2013pseudo, xie2020self, arazo2020pseudo, pham2021meta}, and teacher-student learning \cite{li2017large, matiisen2019teacher, zhang2021teacher}, where a teacher model is used to generate pseudo training targets from a large unlabeled dataset to improve the performance of a student model. More specifically, existing pipelines for source separation include silent source detection \cite{defossez2019demucs}, pseudo label generation and filtering \cite{wang2021semi}, and self-boosting \cite{tzinis2022continual, tzinis2022remixit}. The first one trains a classifier on segments of unlabeled data to detect whether the target source is absent, and if so the segment is used as a clean residual signal for data augmentation. However, as the activation of different musical instruments can be sparse in time, the performance of the classifier may highly dependent on whether the training data is balanced, and when the classifier fails the model may use a mixture segment that also contains the target source as the residual. The second one performs standard pseudo label generation pipeline to extract pseudo labels from unlabeled data, and trains a frame-level separate source activity detector to evaluate whether the extracted pseudo labels are high-quality samples. However, the experiments were only conducted on 16k Hz sample rate data on the vocal-accompaniment separation task, and the configuration of the source activity detector was not described in detail. Moreover, as the separated pseudo-label signals may contain distortions, it may require extra data augmentation and training tricks on the optimization of the detectors in order to allow them work well on a specific pre-trained separator. The last one also generates pseudo labels for data augmentation, while it also continuously improves the teacher by sequentially replacing the teacher by the latest student and double the size of the student model during training. The robustness against the noisy teacher outputs is obtained by generating a large batch of noisy samples with a more diverse data remixing paradigm, together with an unsupervised training objective. However, it was only evaluated on speech enhancement task with 16k Hz sample rate either, and the use of large number of pseudo label samples as well as the increase of the student model size may introduce additional difficulties in the application of the pipeline on large-scale models.

As described above, the data sampling pipeline we use can be viewed as a combination of all existing pipelines -- we use both detected clean segments and generated pseudo-label signals in the unlabeled dataset via a strong pre-trained model. This allows us to not only alleviate the need of an external classifier or source activity detector while still able to detect clean segments within an unlabeled song, but also make use of the pseudo label signals in a similar way as existing pipelines.

\section{Experiment configurations}
\label{sec:config}
\subsection{Supervised Training}

\subsubsection{Data Preprocessing}
\label{sec:data-preprocessing}

Similar to existing works, we use the MUSDB18-HQ dataset \cite{musdb18-hq} for all experiments. During the preparation of the training data, we apply a source activity detector (SAD) to remove the silent regions in the sound tracks and only lead the salient ones for data mixing. Although any existing SAD systems can be directly applied, here we introduce a simple unsupervised energy-based thresholding method to select salient segments from a full track.

Given a unsegmented track and a segment length $L$ measured by duration (e.g., second), we first split the sound track into overlapped segments of length $L$ with an overlap ratio of 50\%. For each segment, we further split it into 10 chunks of length $L/10$ and calculate the energy of them. For silent chunks, we set their energy to a small value $\epsilon=1e-5$. We then calculate an energy threshold of the full sound track by calculating the maximum value of the 15\% quantile of the energy of all chunks and another small value $\gamma=1e-3$. If there are more than 50\% of the chunks in a segment having their energy higher than the threshold, then we define the segment as a salient segment and save it as a valid training data segment. In our training configuration, we set $L=6$ seconds.

\subsubsection{On-the-fly Data Simulation}
\label{sec:data-simulation}

We apply batch-level on-the-fly data simulation by randomly mixing sound tracks from different songs \cite{uhlich2017improving}. Given a training data length $T\leq L$ and the type of target source (e.g., \textit{vocal}, \textit{bass}, \textit{drum} or \textit{other} in MUSDB dataset), we first randomly sample 1 SAD-preprocessed salient segment of length $L$ for all tracks, and then randomly select a chunk of length $T$ (We set $T=3$ seconds by default). For each chunk, we randomly rescale its energy between $[-10, 10]$ dB compared to its original energy. We then randomly drop the each chunk with probability 0.1 to mimic the segments where the target source is inactive. We add up all chunks to form the mixture. To ensure all samples are in the same scale, we rescale both the mixture and the target by the maximum of the maximum absolute value of their samples. 

\subsubsection{Training Objective}

The training objective is defined as the sum of a frequency-domain mean-absolute-error (MAE) loss and a time-domain MAE loss:

\begin{align}
    \mathcal{L}_{obj} = |\vec{S}_r - \bar{\vec{S}}_r|_1 + |\vec{S}_i - \bar{\vec{S}}_i|_1 + |\text{iSTFT}(\vec{S}) - \text{iSTFT}(\bar{\vec{S}})|_1
\end{align}
where $\bar{\vec{S}} \in \mathbb{C}^{F\times T}$ denotes the complex-valued spectrogram of the clean target, subscript $r$ and $l$ denote the real and imaginary parts, respectively, and iSTFT denotes the inverse STFT operator.

\subsubsection{Hyperparameter Configuration}

We set the window size and hop size of STFT to 2048 and 512, respectively, and use a Hanning window. We set the feature dimension $N$ to be 128 in all experiments, and use 12 band and sequence modeling modules with a total of 24 residual BLSTM layers. We set the hidden unit of BLSTM layers to be $2N=256$, the hidden size in the mask estimation MLP to be $4N=512$, and use the hyperbolic tangent function as the nonlinear activation function in the MLP. We use a gated linear unit (GLU) \cite{dauphin2016language} for the output layer of the MLP. The band split bandwidth will be discussed in Section~\ref{sec:bandwith}.

We train individual models for each of the target tracks, which means that we treat the MSS problem as a source extraction problem with only one signal-of-interest. All models are trained for 100 epochs with the Adam optimizer \cite{kingma2014adam} with an initial learning rate of $1e-3$, and each epoch contains 10000 batches of samples with a batch size of 2 and number of GPUs of 8. The learning rate is decayed by 0.98 for every two epochs, and gradient clipping by a maximum gradient norm of 5 is applied. Early stopping is applied when the best validation is not found in 10 consecutive epochs. 

\subsection{Semi-supervised Finetuning}

We use another private dataset of 1750 songs for the semi-supervised finetuning stage. The data sampling process in the on-the-fly simulation pipeline follows the one we described in Section~\ref{sec:sampling}, where we set the MUSDB18-HQ dataset as $\mathcal{L}$ and the private dataset as $\mathcal{U}$. The private dataset is also passed to the source activity detector to remove silent segments before training. We set the initial learning rate for the finetuning stage to $1e-4$, and all other configurations are kept identical to the supervised training stage.

\subsection{Evaluation}

During the evaluation phase, we split the full song into chunks of length $T$ and hop size of $P\leq T$, and perform zero-padding of length $L-P$ at the beginning and the end. All chunks are then processed by the trained model, and overlap-add is applied to all separated outputs to form the final output with the original duration. We set $P=0.5$ seconds by default and discuss the effect of different $P$ in Section~\ref{sec:hop}. We report the model performance on both MUSDB18-HQ and MUSDB18 dataset \cite{musdb18}.

\subsection{Metrics}

We evaluate the models with two metrics:
\begin{enumerate}
    \item \textit{uSDR}: uSDR corresponds to the modified utterance-level signal-to-distortion ratio metric proposed in \cite{mitsufuji2021music} and used as the default evaluation metric in the Music Demixing (MDX) Challenge 2021. The definition of uSDR is identical to the standard signal-to-noise ratio (SNR). We report the mean across the SDR scores of all songs.
    \item \textit{cSDR}: cSDR corresponds to the chunk-level SDR calculated by the standard SDR metric in \textit{bss\_eval} metrics \cite{vincent2006performance} and served as the default evaluation metric in the Signal Separation Evaluation Campaign (SiSEC) \cite{SiSEC18}. We use the official implementation\footnote{\url{https://github.com/sigsep/sigsep-mus-eval}} which reports the median across the median SDR over all 1 second chunks in each song.
\end{enumerate}

Both metrics are reported on decibel scale.

\section{Results and analysis}
\label{sec:result}
\subsection{Effect of Band Split Bandwidth}
\label{sec:bandwith}

The band split bandwidth $\{G_i\}_{i=1}^K$ needs to be manually defined and may affect the performance. We first select seven different options for $\{G_i\}_{i=1}^K$ and compare the models:
\begin{enumerate}
    \item \textit{V1}: We evenly split the entire spectrogram by a 1k Hz bandwidth (remainders are merged to the last subband). This results in 22 subbands.
    \item \textit{V2}: We split the frequency band below 16k Hz by a 1k Hz bandwidth, the frequency band between 16k Hz and 20k Hz by a 2k Hz bandwidth, and treat the rest as one subband. This results in 19 subbands.
    \item \textit{V3}: We split the frequency band below 8k Hz by a 1k Hz bandwidth, split the frequency band between 8k Hz and 16k Hz by a 2k Hz bandwidth, treat the frequency band between 16k Hz and 20k Hz as one subband, and treat the rest as another subband. This results in 14 subbands.
    \item \textit{V4}: We split the frequency band below 1k Hz by a 100 Hz bandwidth, split the frequency band between 1k Hz and 8k Hz by a 1k Hz bandwidth, split the frequency band between 8k Hz and 16k Hz by a 2k Hz bandwidth, treat the frequency band between 16k Hz and 20k Hz as one subband, and treat the rest as another subband. This results in 23 subbands.
    \item \textit{V5}: We split the frequency band below 1k Hz by a 100 Hz bandwidth, split the frequency band between 1k Hz and 16k Hz by a 1k Hz bandwidth, split the frequency band between 16k Hz and 20k Hz by a 2k Hz bandwidth, and treat the rest as one subband. This results in 28 subbands.
    \item \textit{V6}: We split the frequency band below 1k Hz by a 100 Hz bandwidth, split the frequency band between 1k Hz and 4k Hz by a 500 Hz bandwidth, split the frequency band between 4k Hz and 8k Hz by a 1k Hz bandwidth, split the frequency band between 8k Hz and 16k Hz by a 2k Hz bandwidth, treat the frequency band between 16k Hz and 20k Hz as one subband, and treat the rest as another subband. This results in 26 subbands.
    \item \textit{V7}: We split the frequency band below 1k Hz by a 100 Hz bandwidth, split the frequency band between 1k Hz and 4k Hz by a 250 Hz bandwidth, split the frequency band between 4k Hz and 8k Hz by a 500 Hz bandwidth, split the frequency band between 8k Hz and 16k Hz by a 1k Hz bandwidth, split the frequency band between 16k Hz and 20k Hz by a 2k Hz bandwidth, and treat the rest as one subband. This results in 41 subbands.
\end{enumerate}

Table~\ref{tab:bandwidth} provides the vocal extraction performance evaluated by the uSDR metric across the MUSDB18-HQ test set. We can see that the performance of \textit{V1} to \textit{V3} remain on par, but \textit{V4} provides a significant gain. Given that the main difference between \textit{V3} and \textit{V4} is the split of finer subbands below 1k Hz, it shows that lower frequency bands are important for the model to successfully estimate more accurate spectrograms, and a possible explanation is that the frequency band below 1k Hz typically covers the fundamental frequency and the first few harmonics of the vocal track, which enables the band modeling RNN to better capture the F0 information and to better estimate higher frequency components. As more subbands are split at lower frequency parts from \textit{V4} to \textit{V7}, the performance continues to improve, further showing that a fine-grained band-splitting scheme is essential for BSRNN to get better performance. Moreover, we further perform a small-scale grid search and use the following band split bandwidths for the \textit{bass}, \textit{drum} and \textit{other} tracks in MUSDB18:
\begin{enumerate}
    \item \textit{Bass}: We split the frequency band below 500 Hz by a 50 Hz bandwidth, split the frequency band between 500 Hz and 1k Hz by a 100 Hz bandwidth, split the frequency band between 1k Hz and 4k Hz by a 500 Hz bandwidth, split the frequency band between 4k Hz and 8k Hz by a 1k Hz bandwidth, split the frequency band between 8k Hz and 16k Hz by a 2k Hz bandwidth, and treat the rest as one subband. This results in 30 subbands.
    \item \textit{Drum}: We split the frequency band below 1k Hz by a 50 Hz bandwidth, split the frequency band between 1k Hz and 2k Hz by a 100 Hz bandwidth, split the frequency band between 2k Hz and 4k Hz by a 250 Hz bandwidth, split the frequency band between 4k Hz and 8k Hz by a 500 Hz bandwidth, split the frequency band between 8k Hz and 16k Hz by a 1k Hz bandwidth, and treat the rest as one subband. This results in 55 subbands.
    \item \textit{Other}: We use the same band split scheme as \textit{vocals} (i.e., \textit{V7} above).
\end{enumerate}

We empirically find that different instrument tracks may have their own superior band split schemes than that for \textit{vocals}. Given that different instruments can have different frequency ranges, harmonic patterns and mixing techniques, the observation shows that such a priori knowledge or expert knowledge may play an important role in exploring intrinsic characteristics of different musical instruments. Note that adjusting the band split bandwidths may further improve the model performance, and here we do not perform exhaustive grid search for the sake of simplicity.

\begin{table}[!htbp]
    \centering
    \small
    \caption{Performance on vocal extraction for BSRNN models with different band split bandwidths.}
    \begin{tabular}{c|c|c|c|c|c|c|c|c}
    \toprule
        Bandwidth & V1 & V2 & V3 & V4 & V5 & V6 & V7 \\
        \hline
        uSDR & 8.15 & 8.21 & 8.06 & 9.51 & 9.57 & 9.78 & \textbf{10.04} \\
    \bottomrule
    \end{tabular}
    \label{tab:bandwidth}
\end{table}

\begin{table}[!htbp]
    \centering
    \small
    \caption{Performance on vocal extraction for BSRNN models with different evaluation segment hop sizes.}
    \begin{tabular}{c|c|c|c|c}
    \toprule
        Hop size (s) & 0.5 & 1 & 1.5 & 3 \\
        \hline
        uSDR & \textbf{10.04} & 10.00 & 9.94 & 9.75 \\
    \bottomrule
    \end{tabular}
    \label{tab:hop}
\end{table}

\begingroup
\setlength{\tabcolsep}{4pt}
\begin{table*}[!ht]
    \centering
    \scriptsize
    \caption{Comparison with existing models on MUSDB18-HQ (HQ) and MUSDB18 (nHQ) dataset.}
    \begin{tabular}{c|cc|cc|cc|cc|cc|cc|cc|cc|cc|cc}
    \toprule
        \multirow{3}{*}{Model} & \multicolumn{4}{c|}{Vocals} & \multicolumn{4}{c|}{Bass} & \multicolumn{4}{c|}{Drum} & \multicolumn{4}{c|}{Other} & \multicolumn{4}{c}{All} \\
        \cline{2-21}
         & \multicolumn{2}{c|}{uSDR} & \multicolumn{2}{c|}{cSDR} & \multicolumn{2}{c|}{uSDR} & \multicolumn{2}{c|}{cSDR} & \multicolumn{2}{c|}{uSDR} & \multicolumn{2}{c|}{cSDR} & \multicolumn{2}{c|}{uSDR} & \multicolumn{2}{c|}{cSDR} & \multicolumn{2}{c|}{uSDR} & \multicolumn{2}{c}{cSDR} \\
         \cline{2-21}
         & HQ & nHQ & HQ & nHQ & HQ & nHQ & HQ & nHQ & HQ & nHQ & HQ & nHQ & HQ & nHQ & HQ & nHQ & HQ & nHQ & HQ & nHQ \\
         \hline
         ResUNetDecouple+ \cite{kong2021decoupling} & -- & -- & -- & 8.98 & -- & -- & -- & 6.04 & -- & -- & -- & 6.62 & -- & -- & -- & 5.29 & -- & -- & -- & 6.73 \\
         CWS-PResUNet \cite{liu2021cws} & -- & -- & 8.92 & -- & -- & -- & 5.93 & -- & -- & -- & 6.38 & -- & -- & -- & 5.84 & -- & -- & -- & 6.77 & -- \\
         KUIELab-MDX-Net \cite{kim2021kuielab} & -- & -- & 8.97 & 9.00 & -- & -- & 7.83 & 7.86 & -- & -- & 7.20 & 7.33 & -- & -- & 5.90 & 5.95 & -- & -- & 7.47 & 7.54 \\
         Hybrid Demucs \cite{defossez2021hybrid} & -- & -- & 8.13 & 8.04 & -- & -- & \textbf{8.76} & \textbf{8.67} & -- & -- & 8.24 & 8.58 & -- & -- & 5.59 & 5.59 & -- & -- & 7.68 & 7.72 \\
         \hline
         BSRNN & 10.04 & 9.92 & 10.01 & 10.21 & 6.80 & 6.77 & 7.22 & 7.51 & 8.92 & 8.68 & 9.01 & 8.58 & 6.01 & 5.97 & 6.70 & 6.62 & 7.94 & 7.84 & 8.24 & 8.23 \\
         \quad\quad + finetuning & \textbf{10.47} & \textbf{10.36} & \textbf{10.47} & \textbf{10.53} & 7.20 & 7.17 & 8.16 & 8.30 & \textbf{9.66} & \textbf{9.46} & \textbf{10.15} & \textbf{9.65} & \textbf{6.33} & \textbf{6.27} & \textbf{7.08} & \textbf{7.00} & \textbf{8.42} & \textbf{8.32} & \textbf{8.97} & \textbf{8.87} \\
         \hline
    \bottomrule
    \end{tabular}
    \label{tab:all}
\end{table*}
\endgroup

\subsection{Effect of Evaluation Segment Hop Size}
\label{sec:hop}

Table~\ref{tab:bandwidth} reports model performance with the default evaluation segment hop size of $P=0.5$ seconds. We also report the model performance with different segment hop sizes here. Table~\ref{tab:hop} shows the uSDR scores with models with four different hop sizes and a fixed segment size of $T=3$ seconds. We can see that choosing any $P\leq T$ can result in a performance improvement, and the reason can be because the overlap-add operation smooths the outputs and mitigates the noise or distortion introduced. We can also find that decreasing $P$ from 1.5 to 0.5 does not provide significant gain while linearly increases the processing time with the extra segments. This shows that although we choose $P=0.5$ for the experiments here, in practice one can choose $P=1.5$ for a balance between processing speed and performance.

\subsection{Comparison with State-of-the-art Systems}

Here we compare the BSRNN model with existing state-of-the-art systems on both MUSDB18 and MUSDB18-HQ dataset. We choose the top-ranking systems in the Music Demixing (MDX) Challenge 2021 \cite{mitsufuji2021music} as the baselines. We report the results before and after the semi-supervised finetuning stage for BSRNN for all the four tracks. For the other systems, the best reported numbers found in all available literatures are reported. Table~\ref{tab:all} presents the results on both MUSDB18 and MUSDB18-HQ dataset on both uSDR and cSDR metrics. We can observe that BSRNN trained only on MUSDB18-HQ dataset outperforms all existing systems on \textit{vocal}, \textit{drum} and \textit{other} tracks on both MUSDB18 and MUSDB18-HQ dataset, and performs slightly worse on \textit{bass} track. Possible explanations for this observation are that the energy rescaling process in our data mixing procedure might not be well suited for bass as empirically bass is not as strong as other instruments in a song, and the band-split scheme needs further investigation to better capture low- and mid-frequency range information where bass lies in. Regarding the semi-supervised finetuning stage, we observe that all tracks are able to obtain a performance gain, especially for \textit{bass} and \textit{drum} tracks where we can observe an around 1 dB improvement on the cSDR metric. Compared to existing semi-supervised methods described in Section~\ref{sec:semi}, our self-boosting finetuning pipeline is evaluated on all instrument tracks instead of a vocal or speech only task, and the experiments are also conducted on 44.1k Hz sample rate signals instead of 16k Hz sample rate signals. Moreover, our pipeline does not lead to a gradually increased model size as \cite{tzinis2022remixit}, and does not need external modules that require separate training procedures as \cite{defossez2019demucs, wang2021semi}. This proves that the proposed semi-supervised finetuning stage has the potential to become a more universal pipeline for general source extraction or separation tasks.

\section{Conclusion and future works}
\label{sec:conclusion}
In this paper, we proposed \textit{band-split RNN (BSRNN)}, a model architecture that was designed for music source separation and general high-sample-rate source separation that can take a priori knowledge on the characteristics of the source to be separated into account when determining the model hyperparameters. BSRNN split the complex-valued spectrogram of the input mixture in multiple subbands with differend bandwidths, and performed interleaved sequence-level and band-level processing via recurrent neural networks. We also described a semi-supervised data sampling pipeline for finetuning the model trained on a small-scale labeled dataset on a large-scaled unlabeled dataset. Experiment results on MUSDB18 and MUDSB18-HQ dataset showed that BSRNN can surpass the performance of existing state-of-the-art music source separation systems, and the semi-supervised finetuning pipeline can further improve the performance and the robustness on songs with various genres. Future works include the investigation of better and cleverer ways to incorporate a prior source-specific knowledge into the choice of band-splitting schemes rather than large-scale grid search, and the validation of the model and the pipeline on more types of musical instruments and universal audio extraction and separation tasks. Moreover, the size of the private dataset we used for the semi-supervised finetuning stage is not large compared to prior works \cite{defossez2019demucs}, and we only used lossless data with \textit{.wav} or \textit{.flac} formats and removed the songs with all other lossy compression formats (e.g., \textit{.mp3}). As there exists a wide range of songs encoded with lossy compression formats, adding such data to the pipeline may allow the band-level RNN layer to learn to deal with the signals with a distorted higher frequency component, and further improve the model performance on both the MUSDB18 dataset (with \textit{.mp4} format) and other real-world recordings. The effects of additional data size and data type in both supervised training and semi-supervised finetuning stages are also left for future study.

\bibliographystyle{IEEEbib}
\bibliography{refs}

\end{document}